\documentclass[twocolumn,showpacs,preprintnumbers,amsmath,amssymb,prl]{revtex4}

\usepackage[dvips]{graphicx}
\usepackage{dcolumn}
\usepackage{amssymb}
\usepackage{amsfonts}
\usepackage{bm}

\begin{document}

\preprint{APS}
\title{Probing of the Kondo peak by the impurity charge measurement}
\author{M. I. Katsnelson}
\email{M.Katsnelson@science.ru.nl}
\affiliation{Institute for
Molecules and Materials, Radboud University Nijmegen, Nijmegen
6525 ED, The Netherlands}
\author{E. Kogan}
\email{kogan@quantum.ph.biu.ac.il}
\affiliation{Jack and Pearl Resnick Institute, Physics
Department, Bar Ilan University, Ramat Gan  52900, Israel}
\date{\today}

\begin{abstract}
We consider the real-time dynamics of the Kondo system after the local probe
of the charge state of the magnetic impurity. Using the exactly
solvable infinite-degeneracy Anderson model we find explicitly the
evolution of the impurity charge after the
measurement.
\end{abstract}
 \maketitle

The Kondo effect discovered originally in connection with the
resistivity minimum in dilute magnetic alloys \cite{kondo} turns
out to be one of the most interesting many-body phenomena in
condensed matter physics \cite{forty}. The localized
spin-degenerate level embedded in the sea of conduction electrons
may influence the properties of the system in a dramatic way. The
key feature of the Kondo effect is the formation of the Kondo, or
Abrikosov-Suhl, resonance near the Fermi level \cite{hewson}. This
resonance is of crucial importance for so different fields as the
physics of heavy-fermion systems \cite{hewson,stewart} and the
electron transport through quantum dots \cite{qdexp,qdtheo}.
Direct observations of the Kondo resonance at metal surfaces by
the scanning tunneling microscopy method \cite{STM} enhances the
interest to the problem.

In contrast to our detailed understanding of the equilibrium
properties of the Kondo system, much less is known about it's
real-time dynamics. The time development of the Kondo effect in
quantum dots after the voltage change has been considered in Ref.
\onlinecite{nord}. It was shown in Ref. \onlinecite{katsnelson}
that the probe of the charge state of magnetic impurity leads to
suppression of the Kondo resonance. However, the dynamics of the
latter process has not been investigated in detail. In this work
we study the evolution of the Kondo system after the measurement
of the impurity charge.

We proceed with an exactly solvable large-$N_f$ degenerate
Anderson model \cite{GS,largeN} in the infinite-$U$ limit with the
Hamiltonian
\begin{eqnarray}
H=P\left[\sum_{k\nu}[\epsilon_k c_{k\nu}^{\dagger}c_{k\nu}
+\epsilon_f f_{\nu}^{\dagger}f_{\nu}]\right.\nonumber\\
\left.+\frac{1}{\sqrt{N_f}}\sum_{k\nu}\left(V_kf_{\nu}^{\dagger}c_{k\nu}+V_k^*
c_{k\nu}^{\dagger} f_{\nu}\right)\right]P,
\end{eqnarray}
where $c_{k\nu},f_{\nu}$ are the Fermi operators for the
conduction band and $f$-electrons correspondingly (we consider the
case where the $f$-level lies in the conduction band),
$V/\sqrt{N_f}$ is the hybridization parameter, $\nu=1,2,\dots,N_f$
is the "flavor" index and $P$ is the projection operator into the
space with $n_f=\sum_{\nu}f_{\nu}^{\dagger}f_{\nu}<2$. We consider
the case $N_f\to\infty$. In this limit the ground state is
\cite{GS,largeN}
\begin{eqnarray}
\label{psi}
\psi_G=A\left(|\Omega>+\sum_{k\nu}^{\rm occ}
\alpha_kf^{\dagger}_{\nu}c_{k\nu}|\Omega>\right),
\end{eqnarray}
where $|\Omega>$ is the noninteracting vacuum, i.e. a filled Fermi sea
of band electrons with  Fermi energy $\epsilon_F$.

The wave-function after we probe charge localized at the $f$-level
depends upon the results of  the  measurement. If we have found
the hole the wave-function directly after the measurement is
simply
\begin{eqnarray}
\label{g1}
\psi(0)=|\Omega>.
\end{eqnarray}
 Notice that in the accepted approximation
the evolution of the system after the measurement probing hole at the $f$-level
coincides with the model of the initially non-interacting band electrons
where we suddenly switched on hybridization with the $f$-level.

In this limit we have, actually, a single-particle problem with
the effective basis   $|0>=|\Omega>$ and
$|k>=f^{\dagger}_{\nu}c_{k\nu}|\Omega>$ \cite{GS}. The matrix
elements of the Hamiltonian in this basis are
\begin{eqnarray}
\label{ham}
H|0>&=&-\epsilon_f|0>+\sum_{k<k_F} V_k|k> \nonumber\\
H|k>&=&-\epsilon_k|k>+V_k^*|0>,
\end{eqnarray}
describing a discreet level with $E=-\epsilon_f$ embedded into
continuum, presented by a band of a finite width with a bottom at
$E=-E_F$. It is the finite density of states at the bottom of thus
appearing band that is responsible for the formation of
logarithmic divergences and, consequently, the Kondo energy scale
in the problem.

The wave-function  can be presented as
\begin{eqnarray}
\label{ab}
\psi(t)=a(t)|f>+\sum_{k<k_F} b(k,t)|k>,
\end{eqnarray}
with the initial conditions
$a(0)=1$, $b(k,0)=0$.
For the amplitude to find electron at the $f$-level,
straightforward algebra gives
\begin{eqnarray}
\label{int}
a(t)=-\frac{1}{2\pi i}\int_{-\infty}^{\infty}d\omega
\frac{e^{-i\omega t}}
{\omega+\epsilon_f-\Sigma(\omega)}.
\end{eqnarray}
where
\begin{eqnarray}
\label{z}
\Sigma(\omega)=\sum_{k<k_F} \frac{|V_k|^2}{\omega+\epsilon_k}.
\end{eqnarray}
The transition from the discrete  level in the electron band is done by shifting
integration contour by an infinitesimal value $+is$. Thus instead of $N$ poles
at the real axis, corresponding to discrete electron levels we obtain two
branch points: bottom of the
band $\omega=D$ and the Fermi energy $\omega=E_F=0$.

In Eq.(\ref{int}) the integration contour can be closed in the
lower half-plane, so the integral is determined by the
singularities of the integrand in that half-plane. We can take the cuts
along the line $\omega=-iy$, $\infty>y>0$ and $\omega=D-iy$, $\infty>y>0$.
\begin{figure}
\includegraphics[angle=0,width=0.45\textwidth]{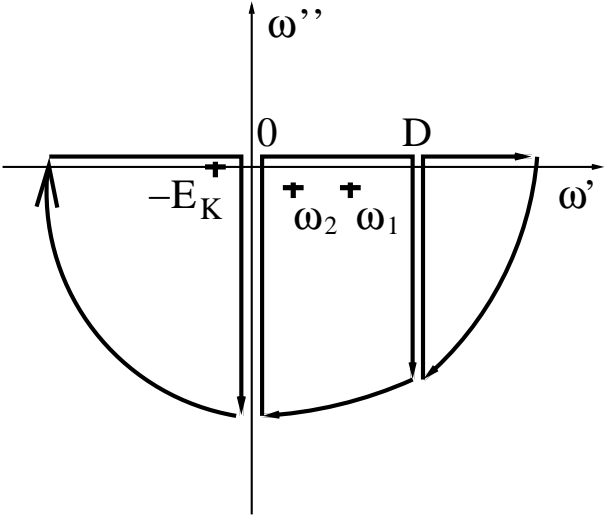}
\caption{\label{fig:cont} Contour used to evaluate integral in Eq. (\ref{int}).}
\end{figure}
Thus
\begin{eqnarray}
\label{ft}
a(t)&=&\sum_{poles,j}e^{-i\omega_jt}R_j+I_{cut},
\end{eqnarray}
where $j$ labels  the poles of the function
$(\omega+\epsilon_f-\Sigma(\omega))^{-1}$ given by the solutions of the equation
\begin{eqnarray}
\label{pole}
\omega_j=-\epsilon_f+\sum_{k<k_F} \frac{|V_k|^2}{\omega_j+\epsilon_k+is},
\end{eqnarray}
and the corresponding residues are
\begin{eqnarray}
R_j=\left[1+\sum_{k<k_F}\alpha_k(\omega_j)\right]^{-1},
\end{eqnarray}
where $\alpha_k(\omega)=V_k^*/(\omega+\epsilon_k)$.
Notice, that for large $t$  only the contribution of
the real pole $\omega=-E_K$ survives, to give
\begin{eqnarray}
\label{asimp}
a(t)=\left[1+\sum_{k<k_F}|\alpha_k(-E_K)|^2\right]^{-1}e^{iE_Kt}.
\end{eqnarray}

Now we present an alternative point of view at the process of
tunneling into continuum, which will clarify the meaning of Eq.
(\ref{int}) and, especially, of Eq. (\ref{asimp}). The eigenstates
of the Hamiltonian (\ref{ham})  can be easily found:
\begin{eqnarray}
\psi_{E}=\frac{|0>+\alpha_k(E)|k>}{\sqrt{1+\sum_{k<k_F}|\alpha_k(E)|^2}},
\end{eqnarray}
and the relevant energies are the solutions of the equation
\begin{eqnarray}
\label{en}
E=-\epsilon_f+\sum_{k<k_F} \frac{|V_k|^2}{E+\epsilon_k}.
\end{eqnarray}
Equation (\ref{en}) being superficially similar to Eq. (\ref{pole}) is totally
different. The  former has only real solutions, their number
beings equal to the number of electron states below the Fermi level plus
one, the latter gives three poles (see below), two of which are
complex, and the third real coincides with the ground state energy  $E=-E_K$.
Notice that in the ground state the $f$-level hole occupation number
is
\begin{eqnarray}
\bar{n}_f=\left[1+\sum_{k<k_F}|\alpha_k|^2(-E_K)\right]^{-1}=1-n_f.
\end{eqnarray}

The wave-function right after the measurement can be presented as
\begin{eqnarray}
\psi(0)=\sum_E\left[1+\sum_{k<k_F}|\alpha_k(E)|^2\right]^{-1/2}\psi_E.
\end{eqnarray}
Hence at any moment $t$ after the measurement
the wavefunction is
\begin{eqnarray}
\label{psit}
\psi(t)=\sum_E\left[1+\sum_{k<k_F}|\alpha_k(E)|^2\right]^{-1/2}\psi_Ee^{-iEt},
\end{eqnarray}
and the amplitude to find a hole at
the level $f$ is
\begin{eqnarray}
a(t)=\sum_E\left[1+\sum_{k<k_F}|\alpha_k(E)|^2\right]^{-1}e^{-iEt},
\end{eqnarray}
which exactly corresponds to the integral from Eq. (\ref{int}), calculated by
the residues method. For $t\to\infty$ due to dephasing of the
contributions from all the states save the bound state we obtain
\begin{eqnarray}
\label{asimp2} a(t)=\bar{n}_fe^{iE_Kt},
\end{eqnarray}
coinciding with Eq. (\ref{asimp}).

For a more detailed analysis of Eq. (\ref{ft}) we
additionally specify our model, assuming  $V_k=V=$ const and
flat-band density of bare itinerant-electron states
$\rho=\rho_0$. In this model we get
\begin{eqnarray}
\Sigma(\omega)=\Delta\ln\left(\frac{\omega}{\omega-D}\right),
\end{eqnarray}
where $\Delta=|V|^2\rho_0$; the imaginary part of the logarithm is
equal to $-\pi$ at the real axis between the branch points to agree with
Eq.(\ref{z}) at the real axis. In this case there exist two complex
poles.

Eq.(\ref{ft}) being valid independently of the strength of the
hybridization, we however limit our analysis to the Kondo regime
$\Delta\ll -\epsilon_f$, where we get
\begin{eqnarray}
E_K&=&De^{-\left|\epsilon_f\right|/\Delta}\nonumber\\
\bar{n}_f&=&\frac{E_K}{\Delta}\ll 1.
\end{eqnarray}
One complex pole $\omega_1=E_f=-\epsilon_f-i\pi\Delta$, with the
residue approximately equal to $1$, is responsible for the
traditional ``Fermi golden rule'' decay processes. The second
pole $\omega_2=E_K-i\pi\Delta$ with the residue equal to
$-\bar{n}_f$ is the complex "mirror" of the Kondo pole. As can be
shown, in the regime we consider, the contribution from the cuts
is negligible for small and large $t$. We may hope the at the
semi-quantitative level it can be neglected at all $t$, though,
strictly speaking, the last claim should be substantiated by
numerical calculations. Thus
\begin{eqnarray}
\label{ft1}
a(t)&=&\bar{n}_f\left(e^{iE_Kt}-e^{-iE_Kt-\pi\Delta t}\right)
 +e^{i\epsilon_ft-\pi\Delta t}.
\end{eqnarray}

Due to an entanglement between the localized electron and the
Fermi sea the local probe of the former should create the
``decoherence wave'' \cite{decohwave} in the conduction electron
subsystem which will disturb the spin and charge distribution in
the latter. Let us calculate the distribution of conduction
electrons around the impurity after the measurement. The
probability density to find the hole at a distance $r$ from the
impurity is given by the expression
\begin{eqnarray}
\label{rho}
\rho_c(r,t)=\left|\sum_{k<k_F} V_ke^{ikr}\frac{1}{2\pi}\int_{-\infty}^{\infty}
\frac{d\omega}{ \omega+\epsilon_k}
\frac{e^{-i\omega t}}{\Sigma(\omega)-\omega-\epsilon_f}\right|^2.
\end{eqnarray}
The asymptotic of Eq.(\ref{rho}) for large $t$ is obvious: the
main contribution comes from the real pole $\omega=-E_K$, and we
obtain
\begin{eqnarray}
\label{rho3}
\rho_c(r)=\left|\bar{n}_f\sum_{k}\frac{V_ke^{ikr}}{E_K-\epsilon_k}\right|^2.
\end{eqnarray}
 Notice that the
coordinate dependence of the hole density of states long after the
measurement is the same as in the ground state; only the
coefficient is smaller by additional factor $\bar{n}_f$ which is
just the suppression factor for the spectral weight of the Kondo
resonance \cite{katsnelson}.

If there exists intermediate asymptotic of exponential decay of
$a(t)$, it is described by the complex pole $E_f$. For these
values of time  the probability density is determined by the
virtually bound state
\begin{eqnarray}
\label{rho33}
\rho_c(r,t)\sim\left|\sum_{k}
\frac{V_ke^{ikr}}{E_f-\epsilon_k}\right|^2e^{-2\pi\Delta t}.
\end{eqnarray}
This asymptotic holds for $\bar{n}_f \ll e^{-\pi\Delta t} \ll 1.$

Assuming for simplicity an isotropic hybridization $V$ and the
dispersion law, we can fulfill angular integration in
Eqs.(\ref{rho3}) and (\ref{rho33}). In the perturbative regime we
consider, $E_K$ is very close to $E_F$. Due to the logarithmic
divergence of the integral in Eq.(\ref{rho3}) the main
contribution comes from the vicinity of the upper limit, and we
obtain
\begin{eqnarray}
\label{rho333} \rho_c(r)=
\frac{\sin^2(k_Fr)}{(k_Fr)^2}\bar{n}_f^{2} \left( \frac{E_K
-\epsilon_f}{V_{k_F}}\right)^2.
\end{eqnarray}

Now we consider another measurement result. Let the first
measurement has found the localized electron sitting at the
$f$-level. The wave-function directly after such measurement
according to the von Neumann ``wave-function collapse'' postulate
\cite{neumann} is again given by Eq. (\ref{ab}) with the initial
condition
\begin{eqnarray}
\label{psi2}
\psi(0)\sim\frac{1}{\sqrt{N}}\sum_{k\nu}^{\rm occ}
\alpha_kf^{\dagger}_{\nu}c_{k\nu}|\Omega>.
\end{eqnarray}
After simple algebra we obtain
\begin{eqnarray}
a(t)=\frac{C}{2\pi i}\int_{-\infty}^{\infty}d\omega
\frac{e^{-i\omega t}}
{\Sigma(\omega)-\omega-\epsilon_f}Y(\omega),
\end{eqnarray}
where
\begin{eqnarray}
Y(\omega)=-\sum_{k}\frac{|V_{k}|^2}
{(\omega+\epsilon_k+is)(E_K-\epsilon_k)}.
\end{eqnarray}
Again closing the contour of the integration and ignoring the
contribution from the branch cuts we find  (in the perturbative
regime)
\begin{eqnarray}
\label{ft2}
a(t)=\sqrt{\frac{\bar{n}_f}{1-\bar{n}_f}}\left[(1-\bar{n}_f)e^{iE_Kt}
\right.\nonumber\\
\left.+\bar{n}_fY(-E_K)
e^{-iE_Kt-\pi\Delta t}
+Y(\epsilon_f)e^{i\epsilon _ft-\pi\Delta t}\right].
\end{eqnarray}
>From the orthogonality of eigen-functions corresponding to $-E_K$ or
$\epsilon_f$ to the wave-function corresponding to $E_K$ we obtain
\begin{eqnarray}
Y(-E_K)=Y(\epsilon_f)=-1.
\end{eqnarray}
Finally, in the Kondo regime ($\bar{n}_f\ll 1$)  one has
\begin{eqnarray}
a(t)=\sqrt{\bar{n}_f}\left[e^{iE_Kt}
-e^{i\epsilon _ft-\pi\Delta t}\right] .
\end{eqnarray}
Thus for $t\to\infty$
\begin{eqnarray}
a(t)=\sqrt{\bar{n}_f}e^{iE_Kt}.
\end{eqnarray}
This result can be understood in a simple way. Since in the limit
under consideration $\Delta\ll -\epsilon_f$ the (hole) occupation
number of the $f$-level is very small, the measurement with the
result $\bar{n}_f=0$ produces negligible disturbance of the
wave-function of the ground state. This is in sharp contrast with
the measurement which results in $n_f=1$. After that measurement,
as we can see from Eq.(\ref{asimp}), the asymptotic value of $n_f$
changes drastically.

We have analyzed two kinds of processes. The process of
measurement instantly reduces the ground-state  wave function,
with the probability to find a hole at the $f$-level equal to
$\bar{n}_f$) to the form (\ref{g1}), with the probability to find
a hole at the $f$-level equal to $1$ (we'll consider here only one
possible measurement result). After the measurements starts the
process of quantum evolution with the characteristic time scale
$1/\Delta$, which ends up in the state with the probability to
find a hole at the $f$-level equal to $\bar{n}_f^2$. Notice that
we obtained seemingly paradoxical  result. On the physical grounds
one should expect the disappearance of the measurement effects at
large times, and hence the return of the hole-occupation number to
$\bar{n}_f$. The way to solve  the paradox is clear when we look
at Eq. (\ref{psit}). Together with the evolution described by this
equation there exists a third kind of a processes: thermalisation
of the hole as the result of the excitation of electron-hole
pairs. The consideration of such processes demands considering
instead of the model $N_f\to \infty$ the model with large but
finite $N_f$ and taking into account terms of the order of
$1/N_f$. This will be the subject of a separate publication, but
already now we can say that in such model the time scale of this
thermalisation will be much larger than the time-scales of the
evolution described by Eq. (\ref{ft1}) and determined by the
inverse Kondo temperature. Thus even in such model, our results,
say Eq. (\ref{asimp2}) will be still valid, only it will present
an intermediate asymptotic. Notice that what is said in the
conclusion is the time-representation of the phenomenon of the
density of states description of the Kondo resonance. Whereas in
the infinite-$N_f$ limit investigated by us the Kondo peak has a
zero width, for finite $N_f$ it is a Lorentzian with both the
distance from the Fermi energy and the width of order of the Kondo
temperature (and for purely spin case $N_f =2$ the center of the
resonance just coincides with the Fermi energy)
\cite{hewson,largeN}.

To conclude, we have presented the analytical solution of the
problem of real time charge dynamics in quantum impurity system
for the exactly solvable infinite-$N_f$ limit. This limit is
sufficient to describe the effects of decoherence on the Kondo
resonance. On the other hand, the process of recoherence, that is,
return to the ground state after the measurement, requires the
consideration of the higher-order processes in $1/N_f$.

\end{document}